\documentclass[12pt,preprint]{aastex}

\slugcomment{}

\shorttitle{Planetary companion around a nearby young star}
\shortauthors{J. Setiawan et al.}

\begin{document}

\title{Evidence for a planetary companion around a nearby young star}


\author{J. Setiawan, P. Weise, Th. Henning, R. Launhardt, A. M\"uller}
\affil{Max-Planck-Institut f\"ur Astronomie, K\"onigstuhl 17, 69117 Heidelberg, Germany}
\email{setiawan@mpia-hd.mpg.de}

\author{J. Rodmann}
\affil{(European Space Agency, ESTEC/SCI-SA, Keplerlaan 1, 2201 AZ Noordwijk, The Netherlands)}





\begin{abstract}
We report evidence for a planetary companion around the nearby young star \mbox{HD 70573}. 
The star is a G type dwarf located at a distance of 46 pc with age estimation between 20 and 300 Myrs. 
We carried out spectroscopic observations of this star with FEROS at the 2.2~m MPG/ESO telescope at La Silla.
Our spectroscopic analysis yields a spectral type of G1-1.5V and an age of about 100 Myrs.
Variations in stellar radial velocity of \mbox{HD 70573} have been monitored since December 2003 until January 2007. 
The velocity accuracy of FEROS within this period is about 10 m/s.
\mbox{HD 70573} shows a radial velocity variation with a period of 852($\pm$12) days and a semi-amplitude of 149($\pm$6) m/s. 
The period of this variation is significantly longer than its rotational period, 
which is 3.3 days. Based on the analysis of the Ca II K emission line, H$\alpha$ and $T_\mathrm{eff}$ variation as 
stellar activity indicators as well as the lack of a correlation between the bisector velocity span and 
the radial velocity, we can exclude the rotational modulation and non-radial pulsations as the source of 
the long-period radial velocity variation. 
Thus, the presence of a low-mass companion around the star provides the best explanation 
for the observed radial velocity variation.
Assuming a primary mass $m_{1}= 1.0 \pm 0.1$ M$_{\mathrm{Sun}}$ for the host star, 
we calculated a minimum mass of the companion $m_{2}\sin{i}$ of 6.1 M$_{\mathrm{Jup}}$, 
which lies in the planetary mass regime, and an orbital semi-major axis of 1.76 AU. 
The orbit of the planet has an eccentricity of $e= 0.4$.
The planet discovery around the young star \mbox{HD 70573} gives an important input for the study 
of debris disks around young stars and their relation to the presence of planets.
\end{abstract}





\section{Introduction}

Precise radial velocity (RV) measurements are a well established technique 
in detecting extrasolar planets around non-active stars, 
like solar-type stars with similar masses and ages to our Sun (see e.g., Butler et al. 2006).
This technique has been also applied in the late 1980's for planet 
searches around cool evolved stars (Cochran \& Hatzes 1989). 
However, the number of extrasolar planets around such non solar-type stars 
is still very small compared to planets around solar-like stars. 
The situation for young stars is similar, 
where practically no convincing case is known so far. 
Planet detections around young and active stars are indeed much 
more difficult than those around evolved and quiet solar-like stars. 

Many young stars possess high levels of stellar activity 
and are also known as fast rotators. Spectroscopically this is indicated  
by strong line broadening and the presence of emission line features, 
in particular H$\alpha$ ($\lambda$6536 \AA), Ca II H ($\lambda$3967 \AA) and K ($\lambda$3934 \AA).
Within the same spectral class the stellar activity of young stars is considerably higher 
than for older stars. The rotational velocity of F-, G- and K-type young stars can be as 
high as a few hundreds km/s which can be observed by strong line broadening.  
This makes precise RV measurements very difficult. 
Intrinsic stellar activity, like non-radial pulsations and rotational modulation, 
manifests itself in RV variation. 
In order to distinguish the sources of RV variation in active stars, the stellar spectra have to 
be investigated carefully, for instance, via the bisector analysis (e.g., Hatzes~1996) 
and stellar activity indicators, like Ca II H \& K emission lines and variation in H$\alpha$ line, to avoid a misinterpretation of the observed RV variation. 
This kind of analysis is indispensable for planet searches around active young stars. 

The search for young planetary systems by the RV technique is indeed limited to 
young stars which do not show a high activity level. Such a high stellar activity affects 
the accuracy of the RV method, like in stars with high rotational velocity ($v\sin{i}>20$ km/s).
Nevertheless, in comparison to other young planet search methods, 
like the direct imaging techniques, the RV method is more sensitive to 
planetary companions with closer orbits, i.e., less than 10~AU to the parent stars.
A further advantage compared to direct imaging is, that the RV method is not strongly 
limited by distance. It can be applied to planet searches in nearby young moving groups (30--70 pc) 
and star-forming regions at $>$\,100\,pc (e.g., the Taurus-Auriga region at 140 pc), 
for which direct imaging methods are not possible.

This work reports the discovery of a planetary companion 
around the nearby young star \mbox{HD 70573}. 
Our RV measurements of \mbox{HD 70573} show a periodic variation on a time scale which is much longer than 
the stellar rotational period. 
This excludes rotational modulation as the source of RV variation.
We will show that the bisector technique allows us to distinguish  
intrinsic stellar activity (non-radial pulsations or 
stellar rotational modulation due to starspots) from variability due to 
companions. 
By measuring the bisector velocity spans we detected rotational modulation 
in other young stars of our sample (Setiawan et al., in preparation). 
The planet detection around \mbox{HD 70573} is concluded by the lack of the 
correlation between the observed RVs and stellar activity indicators (Sect.~4).

\section{HD 70573: A nearby young star}
\mbox{HD 70573} was identified by Jeffries (1995) as a Lithium rich star. 
He predicted an age of this star to be substantially younger than 300 Myrs. 
In a study of young stellar kinematic groups by Montes et al. (2001a), \mbox{HD 70573} 
has been classified as a member of the Local Association (Pleiades moving group) with an age range between 
20 and 150 Myrs. Later, Lop\'ez-Santiago et al.~(2006) classified \mbox{HD 70573} as a member of the Hercules-Lyra association, 
a group of stars comoving in space towards the constellation of Hercules. 
This moving group has an estimated age of $\sim$200 Myrs. 
By comparing the equivalent width of Li $\lambda$6708 \AA~ versus the spectral 
type diagram (Fig.~2 in Montes et al. 2001b), we derived an age within the Pleiades age regime (78--125 Myrs). 

 \begin{table}
 \begin{center}
 \caption{Stellar parameters of \mbox{HD 70573}.\label{tbl-1}}
   \begin{tabular}{lll}
 \tableline
 \tableline 
 Spectral type      &  G1-1.5V		&     \\
 $M_{V}$            &  0.4 		& mag \\
 distance           &  45.7 		& pc  \\
 $m$ 		    &  1.0   $\pm$0.1	& M$_{\sun}$ \\
 $T_{\mathrm{eff}}$ &  5737  $\pm$70	& K    \\
 $[Fe/H]$           &  -0.18 $\pm$0.2	& [Fe/H]$_{\sun}$ \\
 $\log{g}$	    &  4.59  $\pm$0.1	&      \\
 $EW$(Li)	    &  156   $\pm$20   	& m\AA \\
 Age		    &  78--125 	  	& Myrs \\
 $v \sin{i}$	    &  14.7  $\pm$1.0   & km/s \\
 $P_{\mathrm{rot}}$ &  3.296    	& days \\
 \tableline 
 \tableline
 \end{tabular}
 \end{center}
 \end{table}
 
The stellar parameters of \mbox{HD 70573} are compiled in Table~1. 
We measured the equivalent widths (EW) of neutral and ionized lines as 
described in Gray (1992). By comparing our EW measurements with the EWs of standard stars 
adopted from Cayrel de Strobel (2001) and by using the relation between EWs and temperature 
we derive the spectral type of G1-1.5V for \mbox{HD 70573}.
The stellar parameters $T_{\mathrm{eff}}$, [Fe/H], $\log{g}$ have been calculated 
by using the TGV model (Takeda et al. 2002), which computes the stellar parameters from the EW 
of FeI and Fe II.

The absolute visual magnitude has been calculated from the visual brightness $m_{V}=8.70$ mag 
and the distance $d=$ 45.7 pc (Lop\'ez-Santiago et al. 2006).
Henry et al.~(2005) has measured  photometric variations of \mbox{HD 70573} and found a 
period of 3.296 days, which corresponds to the rotational period of the star.
We measured the projected rotational velocity $v \sin{i}$ from the spectral lines 
by using a cross-correlation method (Benz \& Mayor~1981) with the instrumental 
calibration from Setiawan et al.~(2004). Our measured value (see Table\,1) is slightly higher than the value published by  
Henry et al.~(1995), who derived $v\sin{i}=11$ km/s. 

\section{Observations and results}

We are carrying out a RV survey of a sample of young stars with 
FEROS at the \mbox{2.2~m MPG/ESO} telescope located at ESO La Silla Observatory, Chile. 
The spectrograph has a resolution of \mbox{$R=$ 48\,000} and a wavelength coverage of 
3600--9200 \AA~ (Kaufer \& Pasquini 1998).

\begin{figure}[t]
\includegraphics[width=16.0cm, height=12.0cm]{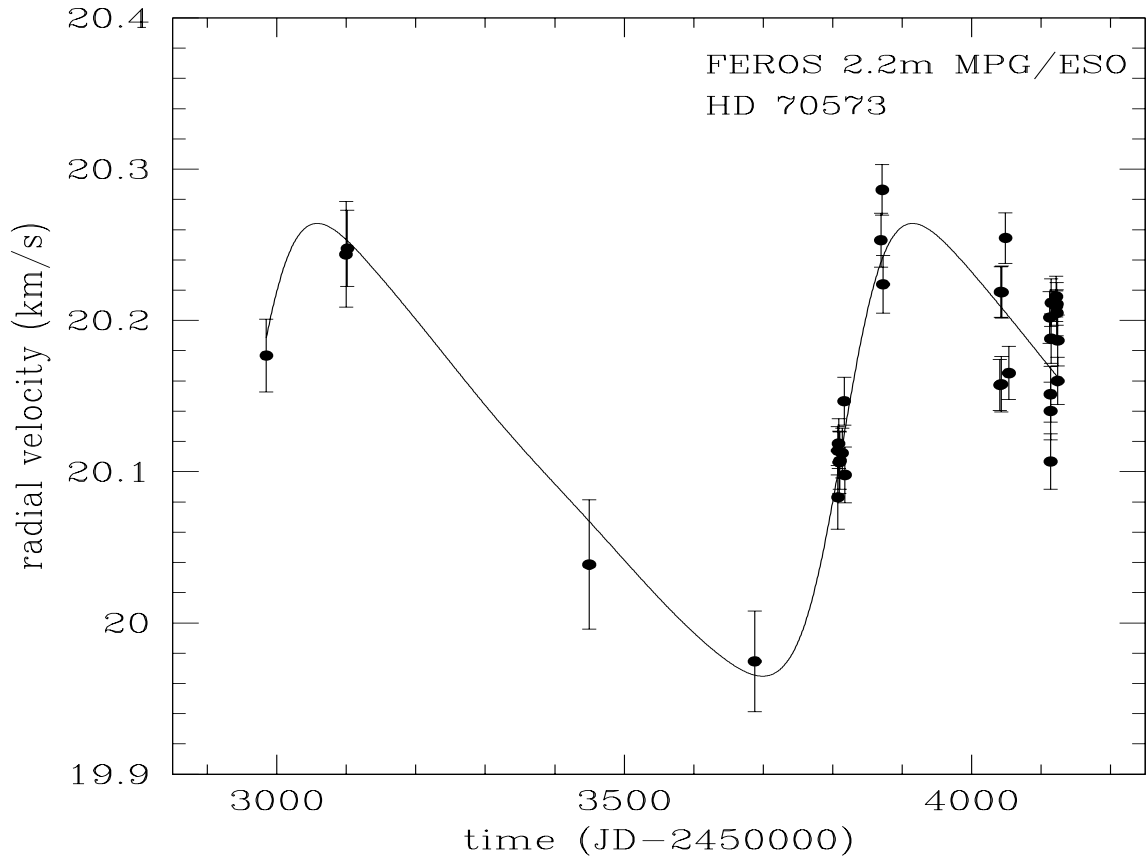}
\caption{RV measurements of \mbox{HD 70573}. We observed a 
long-period RV variation of 852 days and short-period variation 
of few days (see text).}
\end{figure}

The data reduction has been performed by using the online pipeline, which 
produces 39 orders of one-dimensional spectra. 
The RVs have been measured with the simultaneous calibration 
mode of FEROS and a cross-correlation technique (Baranne et al.~1996). 
During the period of three years we obtained a long-term stability of FEROS 
that is about 10 m/s.

RV measurements of \mbox{HD 70573} are shown in Fig.~1. 
We observed a long-term RV variation with a period of 852$\pm$12 days, 
which is much longer than the period of the photometric variability. 
The semi-amplitude of the RV variation is 149$\pm$16 m/s. 
A Lomb-Scargle periodogram (Scargle 1982) of the RVs show the highest peak in the power, 
which corresponds to the long-period RV variation.
On a smaller time scale of several days we also detected short-term RV variations. 
In the Lomb-Scargle periodogram we also found a lower peak in the power, 
which corresponds to a period of $\sim$ 2.6 days. This is comparable to the 
period in the photometric variation detected by Henry et al.~(1995). 
The False Alarm Probability (FAP) of the peaks are $1.1 \times 10^{-3}$ 
for the long-period RV variation and $3.5 \times 10^{-2}$ for the short-period one. 
Additional RV measurements, taken with interval of few hours in several consecutive days, 
may increase the power in the frequency region that corresponds to the period of $\sim$3 days. 

\begin{figure}[t]
\includegraphics[width=16.0cm, height=12.0cm]{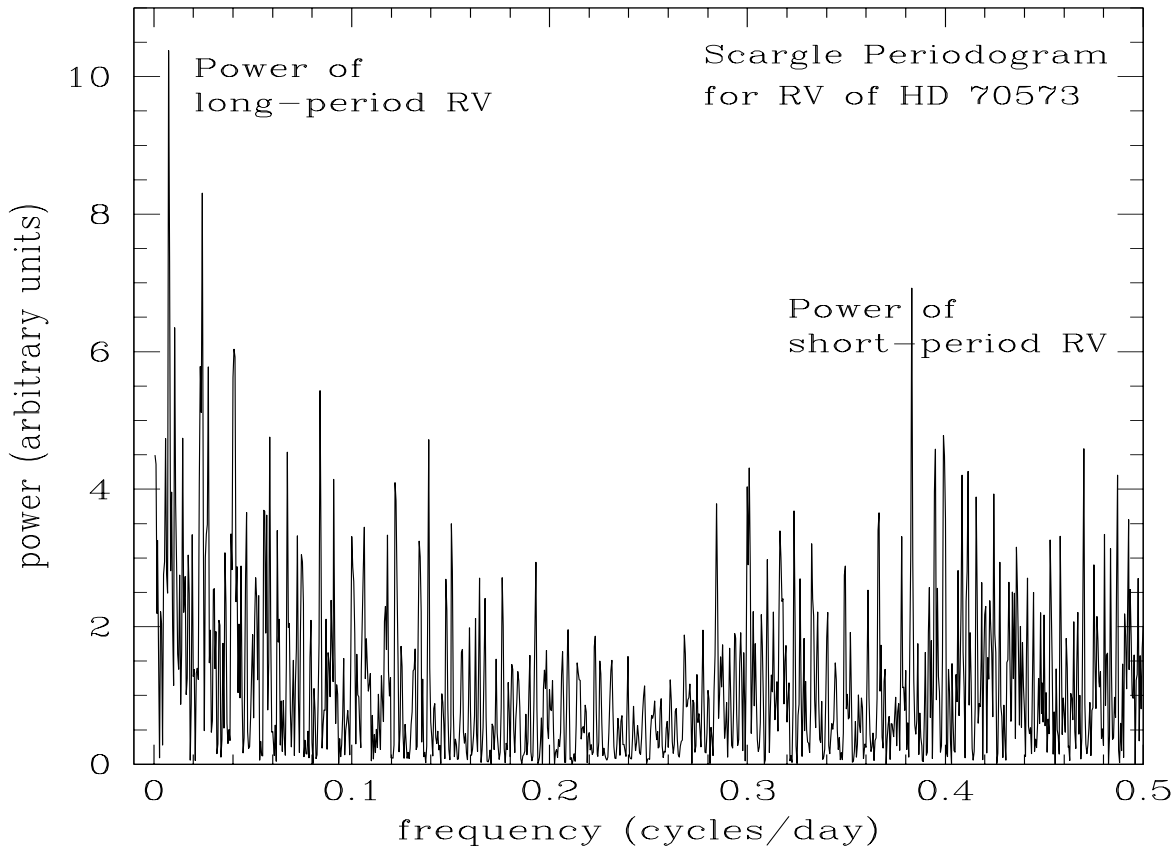}
\caption{Lomb-Scargle Periodogram of the RV variation of HD 70573}
\end{figure}

\begin{figure}[t]
\includegraphics[width=16.0cm, height=12.0cm]{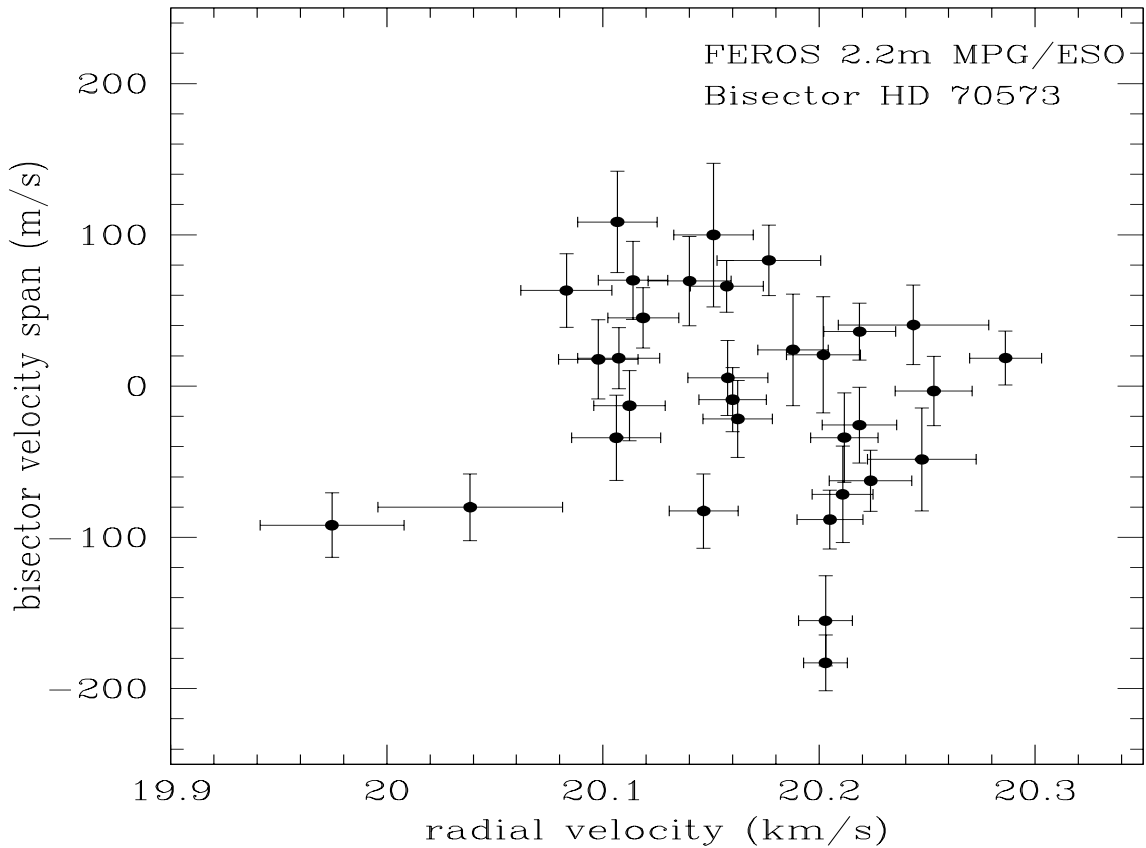}
\caption{Bisector velocity span vs. RV of HD~70573. The figure shows no 
correlation between both quantities. This favors the presence of a low-mass companion 
rather than stellar activity as the source of RV variation.}
\end{figure}



\section{Testing the stellar activity}

As detected in many surveys, young stars show high stellar activity, 
characterized by strong X-ray, H$\alpha$, Ca II H and K emission. 
In addition, they are also known as fast rotators.
For example, large surveys of young stars in star-forming regions such as NGC 2264 (Lamm et al. 2004) 
show that the objects are often fast rotators with periods between 0.2 and 15 days. 
Stellar magnetic activity manifests itself by starspots and granulation, as observed in the Sun. 
Pulsations have also been observed in young stars 
(e.g., Marconi et al. 2000).

To measure the stellar activity of \mbox{HD 70573} we investigated the variation of the 
Ca II K emission line ($\lambda$3934~\AA~) and H$\alpha$. 
We did not use of the Ca II H ($\lambda$3967~\AA)~ to avoid the blend which can be 
caused by the H$\epsilon$ line of the Balmer series. 
Similar to the method used by Santos et al.~(2000), we computed an activity index 
by measuring the intensity of the Ca II K relative to the intensities of 2~\AA~ windows located 
in the blue and red part of the spectra, which are close to the Ca II K region and do not 
have strong absorption features. Our measurements do not show any long period variation which 
might be correlated with the RV variation. 
The relative rms of the $S$-index variation is 4.5\% of the mean value.
In addition, we also measured the equivalent width (EW) variation of the H$\alpha$ line and $T_\mathrm{eff}$ variation 
by using the line-ratio technique (e.g., Catalano et al.~2002) to search for the stellar activity. 
The EW measurements of the H$\alpha$ line give a value of 961$\pm$45 m\AA. 
The rms of 45 m\AA~corresponds to 4.7\% variation in the EW, that is similar to the variation 
observed in the Ca II K emission line. 
We observed a short-term $T_\mathrm{eff}$ variation with a peak-to-peak value of $\sim$220 K 
and a period of few days, which is close to the stellar rotational period. 
This result means an approximately 4\% variation in $T_\mathrm{eff}$ (Table 1) and thus in good agreement 
with other stellar activity indicators.
However, we did not find any long-term periodicity. The equivalent width variation of the H$\alpha$ line 
also does not show any long period variation. 
  
The stellar activity will leave imprints on the spectral line profile.
Another possibility to characterize the stellar activity in the spectra 
is by using the bisector or the bisector velocity span (Hatzes 1996), 
which measures the asymmetry of the spectral line profile. 
Equivalently, the bisector velocity span method can be applied 
to the cross-correlation function used for the RV 
computation (Queloz et al.~2001). 
A correlation between bisector velocity spans and RVs should be expected, 
if the activity is responsible for the RV variation.
In contrast to non-active solar-like stars, the bisector 
velocity spans of active stars are not constant. 
The scatter in the velocity spans may provide information about the activity level 
of the star.
  
 \begin{table}[h]
 \begin{center}
 \caption{Orbital parameters of \mbox{HD 70573} b\label{tbl-2}}
   \begin{tabular}{lll}
 \tableline
 \tableline 
 $P$      	    &  851.8 $\pm$ 11.6		& days  \\
 $K_{1}$            &  148.5 $\pm$ 16.5		& m/s   \\
 $e$                &  0.4   $\pm$ 0.1	    	&       \\
 $\omega$           &  269.6 $\pm$ 14.3		& deg   \\
 $JD_{0}-2450000$   &  2106.54 $\pm$ 25.72	& days  \\
 reduced $\chi^{2}$ &  1.34 			& 	\\
 $O-C$         	    &  18.7 			& m/s\\
 $m_{1}$            &  1.0   $\pm$ 0.1 		& M$_{\sun}$ \\
 $m_{2} sin{i}$     &  6.1   $\pm$ 0.4		& M$_{\mathrm{Jup}}$ \\
 $a$	            &  1.76  $\pm$ 0.05		& AU	\\
 \tableline 
 \tableline
 
 \end{tabular}
 \end{center}
 \end{table}

In \mbox{HD 70573} we found no correlation between the bisector velocity spans and 
RVs (Fig.~3). Thus, based on the results of our analysis of the Ca II K emission lines, 
H$\alpha$, temperature variation and bisector velocity spans as stellar activity indicators
we conclude that the observed long-period RV variation of \mbox{HD 70573} 
is most likely due to the presence of a low-mass (substellar) companion.

\section{Discussion}

We computed an orbital solution for the RV data of \mbox{HD 70573} 
by using a standard Keplerian fit with $\chi^{2}$ minimization.
The orbital parameters are listed in Table~2. \mbox{HD 70573 b} is probably 
the youngest extrasolar planet detected so far with the RV technique (Fig.~4).

\begin{figure}[h]
\includegraphics[width=16.0cm, height=12.0cm]{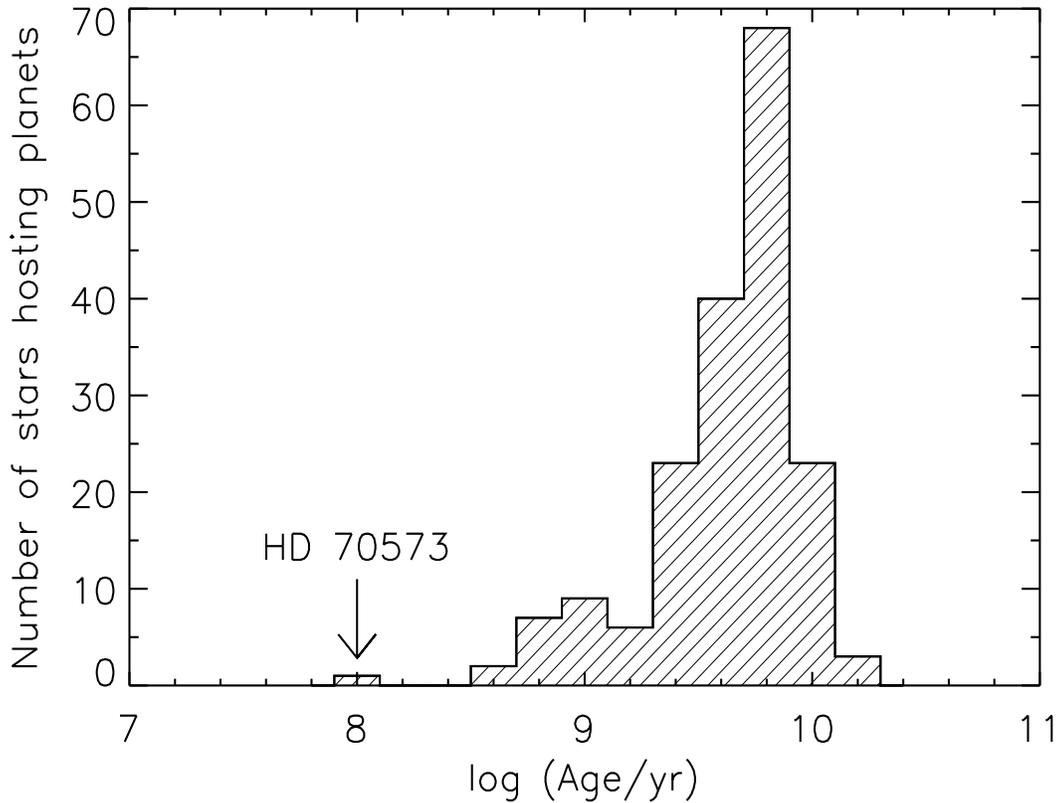}
\caption{A histogram of the ages of exoplanets as of November 2006. \mbox{HD 70573 b} is the youngest planet 
detected so far by the RV method.}
\end{figure}

Planet discoveries around young stars provide important constraints for theories 
of planet formation. An example is the migration process of planets occurring in 
the gas-rich phases of protoplanetary disks. 
The detection of young planets will also allow us to study the relation 
between extrasolar planets and the structure of debris disks (Moro-Mart\'in et al. 2006).
Since \mbox{HD 70573} is part of the young star sample of the SPITZER/FEPS legacy program 
(Meyer et al. 2004), the detection of a planetary companion around this star is of great 
interest for the study of the relation between debris disks and planets. 
With a spectral type of G1-1.5V and an age of only 3--6 \% of the age of the Sun, 
the planetary system around \mbox{HD 70573} could resemble the young Solar system. 

More planet discoveries around young stars will certainly improve our understanding 
of planetary systems in their early evolutionary stages.
Since planet searches around young stars via the RV method are restricted 
to the visual wavelength region and are strongly affected by stellar activity,
other detection techniques like, e.g., NIR direct imaging or astrometry, 
are gaining importance and will most likely soon deliver more discoveries. 
Astrometric measurements with a precision level of few tens of $\mu$as, for example, 
will be able to detect the astrometric signal of the planet around \mbox{HD 70573}, 
which is $\sim$0.23 mas.

Finally, with the detection of a planetary companion around the young star \mbox{HD 70573} 
we have shown, that the RV technique is still potentially profitable for 
the planet search programs.



\acknowledgments

We thank the La Silla Observatory team for the assistance during 
the observations at the 2.2~m MPG/ESO telescope.



{\it Facilities:} \facility{FEROS}, \facility{2.2~m MPG/ESO}.

\clearpage




\clearpage


\begin{thebibliography}{}

\bibitem[Baranne (1996)]{bar96} Baranne, A., Queloz, D., Mayor, M., et al. 1996, A\&AS, 119, 373  

\bibitem[Benz \& Mayor (1981)]{ben81} Benz, W. \& Mayor, M. 1981, A\&A, 93, 235 

\bibitem[Butler et al. 2006]{but06} Butler, R. P., Wright, J. T., Marcy, G. W., et al. 2006, ApJ, 646, 505

\bibitem[Catalano et al.(2002)]{cat02} Catalano, S., Biazzo, K., Frasca, A. et al. 2002, A\&A, 394, 1009  

\bibitem[Cayrel de Strobel et al.(2001)]{cay01} Cayrel de Strobel, G., Soubiran, C., \& Ralite, N. 2001, A\&A, 373, 159  
    
\bibitem[Cochran \& Hatzes (1989)]{cah89} Cochran, W. D. \& Hatzes, A. P. 1989, BAAS, 21, 114  

\bibitem[Gray (1992)]{gra92} Gray, D. F., 1992, The Observation and Analysis of Stellar Photosphere, Cambridge 
University Press 

\bibitem[Hatzes (1996)]{hat96} Hatzes, A. P. 1996, PASP, 108, 839 

\bibitem[Henry et al. (1995)]{hen95} Henry, G. W., Fekel, F. C., \& Hall, D. S. 1995, AJ, 110, 2926 

\bibitem[Jeffries (1995)]{jef95} Jeffries, R. D. 1995, MNRAS, 273, 559

\bibitem[Lamm (2004)]{lam04} Lamm, M., Mundt, R., Bailer-Jones, C. et al. 2004, A\&A, 417, 557

\bibitem[Lop\'ez-Santiago et al.]{lop06} Lop\'ez-Santiago, J., Montes, D., Crespo-Chacon, L. \& Fernandez-Figueroa,
M.J. 2006, ApJ, 643, 1160 

\bibitem[Kaufer \& Pasquini 1998]{kau98} Kaufer, A., \& Pasquini, L. 1998, The Messenger, 95

\bibitem[Marconi et al. 2000]{mar00} Marconi, M., Ripepi, V., Alcal\'a, J.M.  et al. 2000, A\&A, 355, L35

\bibitem[Meyer et al. 2004]{mey04} Meyer, M. R., Hillenbrandt, L. A., Bachman, D. E. et al. 2004, ApJS, 154,
422 

\bibitem[Montes et al. (2001a)]{mon01a} Montes D., Lop\'ez-Santiago, J., Galvez, M. C. et al. 2001, MNRAS, 328, 45

\bibitem[Montes et al. (2001b)]{mon01b} Montes D., Lop\'ez-Santiago, J., Fern\'andez-Figueroa, M. J. et al.
2001, A\&A, 379, 976

\bibitem[Moro-Martin et al. 2006]{mor07} Moro-Mart\'in, A., Carpenter, J. M.,  et al. 2006, Astro-ph/0612242 



\bibitem[Queloz et al. (2001)]{que05} Queloz, D., Henry, G. W., Sivan, J. P., et al. 2001, A\&A, 379, 279

\bibitem[Santos et al. (2000)]{san00} Santos, N.C., Major, M., Naef, D. et al., 2000, A\&A, 361, 265

\bibitem[Scargle J. D.]{sca82} Scargle, J. D. 1982, ApJ, 263, 835

\bibitem[Setiawan et al. (2004)]{set04} Setiawan J., Pasquini, L., da Silva, L. et al., 2004, A\&A, 421, 241

\bibitem[Takeda et al. (2002)]{tak02} Takeda, Y., Ohkubo, M. \& Sadakane, K. 2002, PASJ, 54, 451



 
\end{thebibliography}
\end{document}